\documentclass{jpconf}                  
             
\usepackage{graphicx}
\usepackage{amssymb}
\usepackage{amsmath}

\begin{document}

\title{Loewner equation for  Laplacian growth: A Schwarz-Christoffel-transformation approach}

\author{M. Dur\'an and G.~L.~Vasconcelos}

\address{Laborat\'orio de F\'{\i}sica Te\'orica e Computacional, 
Departamento de F\'{\i}sica, Universidade Federal de Pernambuco,
50670-901, Recife, Brazil.}
\ead{giovani@df.ufpe.br}



\date{\today}

\begin{abstract}
The problem of  Laplacian growth  is considered within the  Loewner-equation framework. A new method  of deriving the Loewner equation  for a large class of growth problems in the half-plane is presented. The method is based on the Schwarz-Christoffel transformation between the so-called `mathematical planes' at two infinitesimally separated times. Our method not only reproduces the correct Loewner evolution for the case of slit-like fingers but also can be extended to treat more general  growth problems. In particular, the Loewner equation for the case  of a bubble growing into the half-plane is presented.

\end{abstract}



\section{Introduction}

The Loewner equation \cite{loewner} 
is an important result in the theory of univalent functions \cite{univalent}  that has found  important applications  in nonlinear dynamics, statistical physics, and conformal field theory  \cite{review1,BB}.  In its most basic formulation, the Loewner equation  is a first-order differential equation for the conformal mapping $g_t(z)$ from a given `physical domain,' consisting of a complex region  $\mathbb{P}$  minus  a curve $\Gamma_t$ emanating  from its boundary, onto a `mathematical domain' represented by $\mathbb{P}$ itself. 
Usually, $\mathbb{P}$  is either the upper  half-plane or the exterior of the unit circle, but recently the Loewner equation for the channel geometry was also considered \cite{poloneses}.
The  Loewner equation depends on a driving function, here called $a(t)$, that is the image of the growing tip under the mapping $g_t(z)$. An important development on the theory of the Loewner equation  was the discovery by Schramm \cite{schramm} that when the driving function $a(t)$ is a Brownian motion the resulting Loewner evolution describes
 the scaling limit of certain statistical mechanics models. This result spurred great interest in the so-called stochastic Loewner equation  \cite{BB}.

Recently, the deterministic Loewner equation  was also used to study the problem of Laplacian fingered growth in both the half-plane and radial geometries \cite{makarov,selander} as well as in  the channel geometry \cite{poloneses}. In this case, the driving function $a(t)$ has to follow a specific time evolution in order to ensure that the finger tip grows along the gradient lines of the corresponding Laplacian field. The idea of using iterated conformal maps to generate aggregates was first deployed  by Hastings and Levitov  \cite{hastings_levitov} in the context of stochastic growth models, such as  diffusion limited aggregation; see, e.g.,  \cite{procaccia} for further developments along those lines. A deterministic version of the Hastings-Levitov model that is closely related to the Loewner-equation approach---albeit not using explicitly such a formalism---was studied in \cite{hastings}.  

In this paper we consider the problem of Laplacian growth within the context of the Loewner evolution, and present a new method of deriving the corresponding Loewner equation for a broad class of growth models in the half-plane.   Our method is based on the Schwarz-Christoffel (SC) transformation between the mathematical planes $g_t$ and $g_{t+\tau}$, where $\tau$ is an infinitesimal time interval. More specifically, the method consists of expanding the integrand of the SC formula in powers of  the appropriate infinitesimal quantity (related to $\tau$) and then performing the integrals up to the leading-order term. Our method  correctly yields the Loewner evolution for the case of slit-like fingers studied before \cite{poloneses,selander}. More importantly, the method is able to handle more general growth problems, so long as the growth rule can be
specified  (in the mathematical plane)  in terms of a polygonal curve, in which case  the Schwarz-Christoffel transformation can be used. An example is given for the case of a bubble growing from the real axis into the upper half-plane.  

\section{Loewner Equation for Slit-like Fingers}
\label{sec:2}

\subsection{The Case of a Single Finger}
\label{sec:2a}

\begin{figure}[t]
\begin{center}
\includegraphics[width=0.6\textwidth]{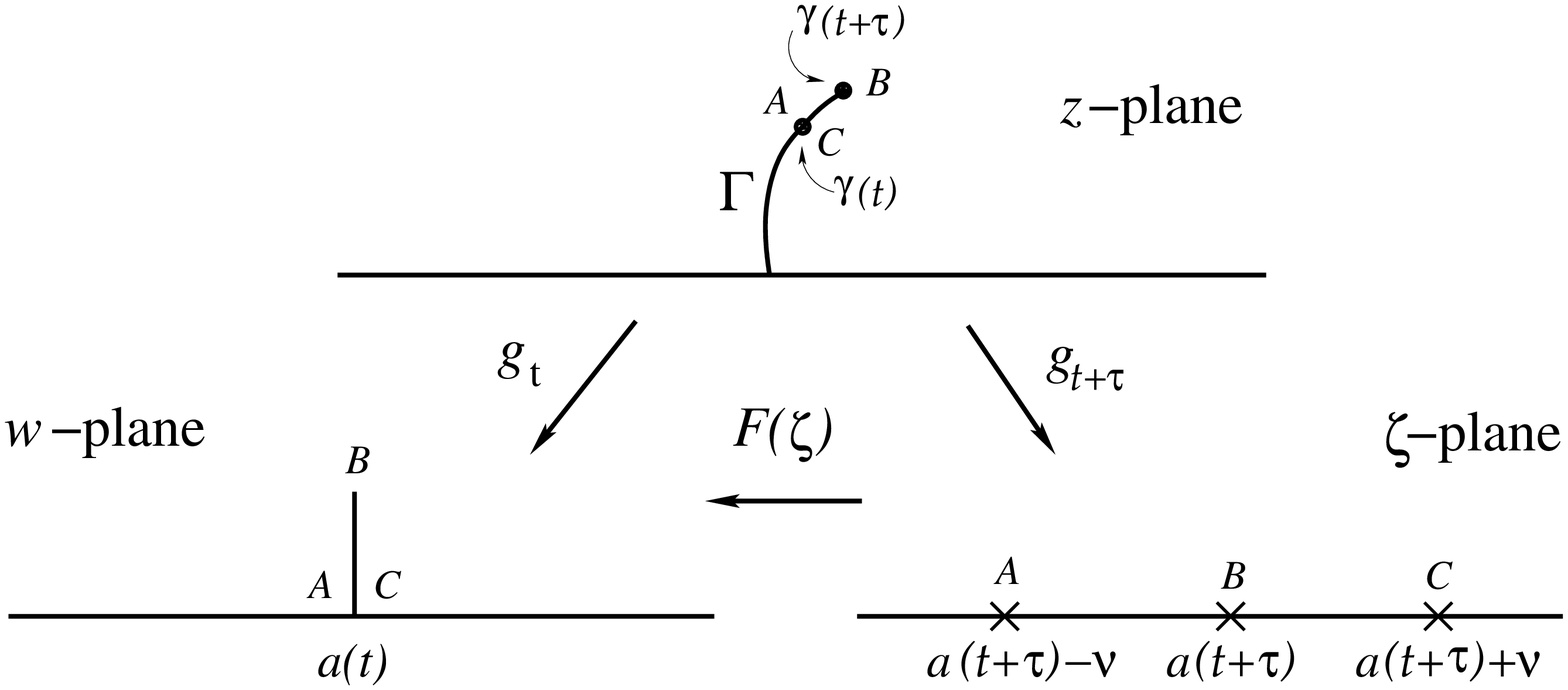}
\end{center}
\caption{The physical $z$-plane and the mathematical $w$- and  $\zeta$-planes at times $t$ and $t+\tau$, respectively, for a single  finger in the upper half-plane.   The mapping $g_t$ maps the curve $\Gamma$ (at time $t$) onto a segment of the real axis on the $w$-plane, whereas the accrued portion of the curve (during time interval $\tau$) is mapped to a vertical slit. The mapping $g_{t}$ is obtained as the composition of $g_{t+\tau}$ and the slit mapping $F$; see text.}
\label{Fig. 1.}
\end{figure}

In order to set the stage for the remainder of the paper, we wish to begin our discussion by considering  the simplest  Loewner evolution, namely, that in which a curve starts from the real axis  at $t=0$ and then grows into the upper half-$z$-plane $\mathbb{H}$, where
\begin{equation*}
 \mathbb{H} = \{ z=x+iy \in \mathbb{C}: y > 0\}  .
\end{equation*}
The curve at time $t$ is denoted by $\Gamma_{t}$ and its growing tip is labeled by $\gamma(t)$.
Now let  $g_t(z)$ be the conformal mapping that maps the  `physical domain,' corresponding to the upper half-$z$-plane minus the curve $\Gamma_t$,  onto the upper half-plane of an auxiliary complex $w$-plane, called the `mathematical plane,' i.e., we have $w=g_t(z)$, where
 \begin{equation}
  g_t: \mathbb{H} \backslash \Gamma_{t} \rightarrow \mathbb{H}  ,
  \end{equation}
with the curve tip $\gamma(t)$ being mapped to a point $a(t)$ on the real axis in the $w$-plane; see Fig.~\ref{Fig. 1.}.  
Furthermore, we consider the growth process to be such that the accrued portion of the curve from $t$ to $t+\tau$, where $\tau$ is an infinitesimal time interval, is mapped under $g_{t}(z)$ to a vertical slit in the mathematical $w$-plane; see Fig.~\ref{Fig. 1.}. The mapping function $g_t(z)$ must also satisfy the initial condition
\begin{equation}
g_{0}(z)=z,
\label{eq:g0}
\end{equation}
since we start with an empty  upper half-plane.
We also impose the so-called hydrodynamic normalization condition at infinity:
\begin{equation}
\label{tres}
 g_t(z) =z + O(\frac{1}{|z|}), \qquad {z \rightarrow \infty} .
\end{equation}
These conditions specify uniquely the mapping function $g_t(z)$.

From a more physical viewpoint, the problem formulated above belong to the class of Laplacian growth models where an interface evolves between two phases driven by 
a scalar field $\phi(x,y;t)$, representing, for example, temperature, pressure, or concentration, depending on the physical problem at hand.  In one phase, initially occupying the entire upper half-plane, the scalar field $\phi$ satisfies the Laplace equation
\begin{equation}
\nabla^{2}\phi=0,
\end{equation}
whereas in the other phase one considers $\phi$=const., say $\phi=0$, with the curve $\Gamma_t$ representing  a finger-like advancing interface between the two phases. (Here the finger is assumed to be infinitesimally thin.) The complex potential  for the problem can then be defined as $w(z,t)=\psi(x,y;t)+i\phi(x,y;t)$, where $\psi$ is the function harmonically conjugated to  $\phi$.
On the boundary of the physical domain, consisting here of the real axis together with the curve $\Gamma_t$, we impose the condition $\phi=0$, whereas at infinity we assume a uniform gradient field, $\vec{\nabla}\phi\approx \hat{y}$, or alternatively,
\begin{equation}
\label{eq:w}
w(z,t) \approx z, \qquad {z \rightarrow \infty} .
\end{equation}
From this point of view, the mapping function $g_{t}(z)$ introduced above corresponds precisely to the complex potential $w(z,t)$ of the problem. In particular, the fact that  in the $w$-plane the curve grows along a vertical line implies that the finger tip grows along gradient lines in the $z$-plane. To specify completely a given physical model, one has also to prescribe the interface velocity, which is usually taken to be proportional to some power $\eta$ of the gradient field: $v\sim|\vec{\nabla}\phi|^{\eta}$.
For most of the problems considered here the specific velocity model  is not relevant, in the sense that the finger shapes will be independent of  the exponent $\eta$, which only affects the time scale of the problem. (However, there are situations, such as the case of competing asymmetrical fingers \cite{poloneses}, where different $\eta$'s may yield different patterns.)

For convenience of notation, we shall represent  the mathematical plane at time $t+\tau$ as the complex $\zeta$-plane 
and so we write $\zeta=g_{t+\tau}(z)$.  Now consider  the mapping
$w=F(\zeta)$, from 
 the upper half-$\zeta$-plane onto the mathematical domain in the $w$-plane; see Fig.~\ref{Fig. 1.}.  The mapping function $g_{t+\tau}(z)$ can then be given in terms of $g_{t}(z)$ as 
\begin{equation}
 g_{t+\tau}=F^{-1}\circ g_{t},
 \label{eq:1}
\end{equation}
 where $F^{-1}$ is the inverse of $F(\zeta)$.  The above relation  governs the time evolution of the function $g_t(z)$ and naturally  leads to the Loewner equation. A standard way  of showing this is to construct the slit mapping
$F(\zeta)$ explicitly, substitute its inverse in (\ref{eq:1}), and then take the limit $\tau\to0$. One then finds the Loewner equation
\begin{equation}
 \dot{g}_t(z) = \frac{d(t)}{g_t(z)-a(t)},
\label{loewner} 
\end{equation}
 where $d(t)$ is the so-called growth factor  which is related to the tip velocity. 
  One can show \cite{poloneses} that $d(t)=|f_{t}^{\prime\prime}(a(t))|^{-\eta/2-1}$, where $f_{t}(z)$ is the inverse of $g_{t}(z)$. Here, however, the specific form of $d(t)$ is not relevant since we can always rescale the time coordinate in (\ref{loewner}) so as to set $d(t)=1$. 
From symmetry, one also gets
 \begin{equation}
\dot{a}(t)=0,
\label{eq:dota}
\end{equation}
so that $a(t)=a_0=$ const., which implies that the tip $\gamma(t)$ simply traces out a vertical line in the $z$-plane.

In more general situations, such as the case of multiple fingers,  the function $F(\zeta)$ can no longer be obtained in closed form and so one has to resort to alternative approaches  to derive the Loewner  evolution. Previous methods \cite{poloneses,selander} consider a series of compositions of the basic one-slit mapping. Here, however,  we will apply a more direct method based on the Schwarz-Christoffel transformation to obtain the Loewner equation for slit-like fingers in the half-plane. In the next section, our method will be extended to treat the case of a growing bubble in the upper half-plane. To illustrate how the method works,   let us first use it to derive the Loewner equation (\ref{loewner}) for a single finger.

We begin by inverting (\ref{eq:1}) so as to write
 \begin{equation}
 g_{t}=F\circ g_{t+\tau}.
 \label{eq:2}
 \end{equation}
The mapping $F(\zeta)$ from the upper half-$\zeta$-plane onto the upper half-$w$-plane with a vertical slit (see Fig.~\ref{Fig. 1.}) is easily found by a direct application of the Schwarz-Christoffel formula \cite{CKP}. One then finds
\begin{equation}
\label{dois}
 g_t = F(g_{t+\tau}) =  \int_{\zeta_0}^{g_{t+\tau}} \frac{\zeta - a(t+\tau)}{\sqrt{(\zeta - a(t+\tau))^2-\nu^2}} d\zeta + a(t),
\end{equation}
where $\zeta_0=a(t+\tau)+\nu$. Note from Fig.~\ref{Fig. 1.} that the parameter $\nu$ is related to the (infinitesimal)  height of the slit in the $w$-plane.
The above integral can be performed exactly, as already mentioned,
but here we take an alternative approach, namely, we first expand the integrand in (\ref{dois}) in powers of $\nu$ and then compute the relevant integrals afterwards. To do that we first rewrite (\ref{dois}) in the form
\begin{equation}
\label{eq:gt}
 g_t =  \int_{\zeta_0}^{g_{t+\tau}} \frac{d\zeta}{\sqrt{1-\nu^2[\zeta - a(t+\tau)]^{-2}}} + a(t).
\end{equation}
After expanding the integrand in powers of $\nu^2$ and performing the relevant
integrals, one obtains that, up to order $\nu^2$, equation (\ref{eq:gt}) becomes
\begin{equation}
g_{t+\tau} -g_t =   \frac{\nu^2/2}{g_{t+\tau}-a(t+\tau)} +  a(t+\tau) - a(t).
\label{eq:gtt}
\end{equation}
Now expanding this equation up to the first order in $\tau$, dividing by $\tau$, and then taking $\tau\to0$, one gets
\begin{equation}
\dot{g}_t  = \frac{d(t)}{g_t(z)-a(t)}+ \dot{a},
\label{eq:dg}
\end{equation}
where
\begin{equation}
d(t) = \lim_{\tau\to0} \frac{\nu^2}{2\tau} .
\label{eq:d}
\end{equation} 
Using the boundary condition  $\lim_{g_{t} \to \infty}\dot{g}_t=0$, which follows from (\ref{tres}), yields precisely the Loewner equation (\ref{loewner}) together with the condition (\ref{eq:dota}). 

\subsection{The Multifinger Case}

\begin{figure}[t]
\centering
\includegraphics[width=0.6\textwidth]{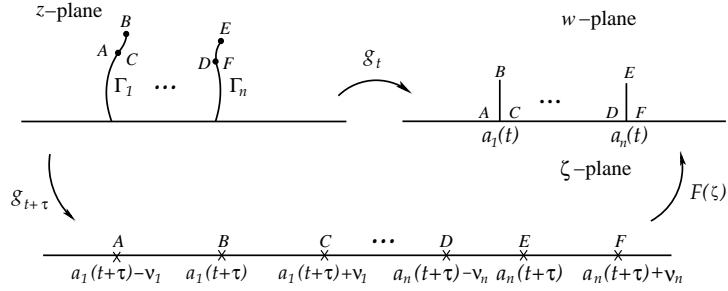}
\caption{The physical and mathematical planes for the case of multiple fingers in the upper half-plane.}
\label{Fig. 2.}
\end{figure}

Here we consider the case of multiples fingers $\Gamma_i$, $i=1,...,n$, growing from the real axis into the upper half-$z$-plane, as shown in Fig.~\ref{Fig. 2.}. As before,  the map $g_t(z)$ maps the physical domain in the $z$-plane onto the upper half-$w$-plane and the tips are required to grow along  gradient lines, so that  the accrued portions of the curves $\Gamma_i$ during an infinitesimal time $\tau$ are mapped under $g_t(z)$ to vertical slits emanating from the real axis; see Fig~\ref{Fig. 2.}.

The mapping $w=F(\zeta)$, from the upper half-$\zeta$-plane to the  upper half-$w$-plane with $n$ vertical slits,
can again be easily obtained from the Schwarz-Christoffel transformation \cite{CKP}: 
\begin{equation}
  g_t = F(g_{t+\tau}) =\int_{\zeta_0}^{g_{t+\tau}} \prod_{i=1}^{n} \frac{\zeta-a_i(t+\tau)}{\sqrt{(\zeta-a_i(t+\tau))^2-\nu_i^2}}\, d\zeta + a_j(t),
\label{eq:8}
\end{equation}
where $\zeta_0= a_j(t+\tau) + \nu_j$ for a given $j$. 
[We remark parenthetically that in writing (\ref{eq:8}) we have assumed,  for simplicity, that the slits in the $w$-plane are mapped under $F^{-1}(w)$ onto symmetrical segments on the real $\zeta$-axis;  see Fig~\ref{Fig. 2.}. Rigorously speaking, this is valid only in the limit that the slit heights become vanishingly small, that is, when $\tau\to0$, which is the relevant limit for us here.]

The  integral in (\ref{eq:8}) cannot be performed exactly for arbitrary $n$, hence in order to obtain the Loewner equation for this case we first need to expand the integrand in powers of the infinitesimal parameters $\nu_i$ and then proceed with the integration. Note, however, that each term in (\ref{eq:8}) is of the same form as that appearing in (\ref{dois}) for the case of a single finger. We can thus build upon our experience with that case to treat the present situation. In particular, we notice that the mixed terms involving different $\nu_i$'s in the expansion of the integrand in (\ref{eq:8}) are all of orders higher than $\nu^2$ and hence need not be considered, for they do not contribute to the final result in the limit $\tau\to0$. Thus, to the extent that the mixed terms can be neglected, we can rewrite (\ref{eq:8}) as
\begin{equation}
  g_t  \approx  \sum_{i=1}^{n} \int_{\zeta_0}^{g_{t+\tau}}\frac{d\zeta}{\sqrt{1-\nu_i^2[\zeta-a_i(t+\tau)]^2}} + a_j(t).
\label{eq:8b}
\end{equation}
Now repeating the exactly same procedure used for the single finger case, see Eqs.~(\ref{eq:gt})-(\ref{eq:dg}), one readily obtains
\begin{equation}
 \dot{g}_t = \sum_{i=1}^n\frac{d_i(t)}{g_t-{a}_i(t)} -  \sum_{ \stackrel{i=1}{i\ne j}}^n\frac{d_i(t)}{{a}_j(t)-{a}_i(t)} + \dot{a}_j(t) ,
 \label{eq:dgt}
\end{equation}
 where
\begin{equation}
d_i(t) = \frac{\nu_i^2}{2\tau} .
\label{eq:di}
\end{equation}
After using the condition  $\lim_{g_{t} \to \infty}\dot{g}_t=0$ in (\ref{eq:dgt}), we get the
Loewner equation for multiple curves
\begin{equation}
 \dot{g}_t = \sum_{i=1}^{n} \frac{d_i(t)}{g_t-{a}_i(t)}, 
\label{eq:11}
\end{equation}
with the time evolution of the points $a_i(t)$ being given by the following system of ordinary differential equations
\begin{equation}
 \dot{a}_i(t) = \sum_{ \stackrel{j=1}{j\ne i}}^n \frac{d_j(t)}{{a}_i(t)-{a}_j(t)} .
 \label{eq:12}
\end{equation}

If the growth factors $d_{i}(t)$ are all the same, we can again rescale the time variable so as to set $d_{i}=1$. In particular, in the case of two symmetrical fingers (i.e.,  $d_1=d_2=1$),  equation (\ref{eq:11}) can be integrated exactly to yield the mapping function $g_t(z)$, from which the finger shapes can be computed analytically \cite{poloneses}. A related exact solution for two fingers was obtained in \cite{kadanoff2}.
An alternative derivation of (\ref{eq:11}) was given elsewhere \cite{selander} using a composition of $n$ single-slit mappings.
Our method, however, is somewhat more direct in the sense that it considers a single mapping with $n$ slits as shown in Fig~\ref{Fig. 2.}. 
 In the next section we will extend our method to include the case of a growing  bubble in the half-plane.  

\section{Loewner Equation for a Bubble in the Half-Plane}
\label{sec:3}

\begin{figure}[t]
\centering
\includegraphics[width=0.6\textwidth]{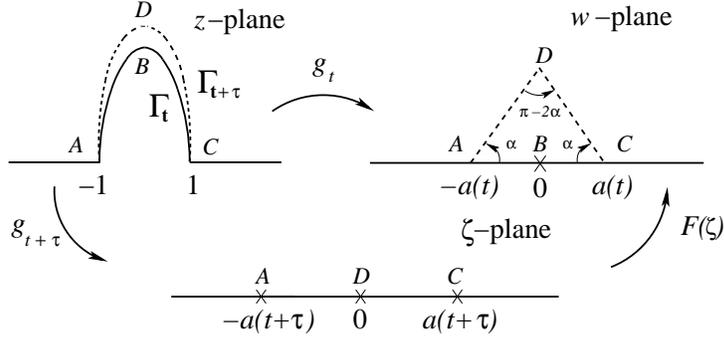}
\caption{The physical and mathematical planes for the case of a growing bubble in the upper half-plane. Here the new interface at time $t+\tau$ is mapped 
under $g_{t}$ to a tent-like shape; see text for details.}
\label{Fig. 4.}
\end{figure}

Here we consider the problem of an interface starting initially from a  segment, say, $[-1,1]$,  along the real axis and then growing into the upper half-$z$-plane, as indicated in Fig.~\ref{Fig. 4.}.  
As before, we denote the interface at time $t$ by $\Gamma_t$ and consider the mapping $g_t(z)$ from the physical  domain in the $z$-plane to the mathematical $w$-plane:
\begin{equation*}
 g_t:\mathbb{H} \backslash \Gamma_t \rightarrow \mathbb{H}  .
\end{equation*}
The mapping function  $g_t(z)$ must also satisfy the hydrodynamic normalization condition (\ref{tres}). The growth dynamics is specified by requiring that each point on the interface grow along gradient lines, but with a growth rate that is maximum at the center (i.e., at the bubble tip) and zero at both `contact points,' $z=\pm 1$, in such a way that  the interface at time $t+\tau$ is mapped under $g_t(z)$ to a tent-like shape, as shown in  Fig.~\ref{Fig. 4.}.  The height of this tent, or alternatively, the angle $\alpha$ formed with the real axis, is linearly related  to  $\tau$ in the limit $\tau \to 0$, as we will see below. 
 Here we consider only the case of symmetric bubbles, where the contact points, $z=\pm 1$, are mapped by $g_{t}(z)$ to the points $w=\pm a(t)$, respectively, and the tip (point B in Fig.~\ref{Fig. 4.}) is mapped to the origin.

Since the domain in the $w$-plane has a polygonal shape, the mapping $w=F(\zeta)$ can once again be obtained from the Schwarz-Christoffel transformation. In this case we have
\begin{equation}
 g_t = F(g_{t+\tau})= \int_{a(t+\tau)}^{g_{t+\tau}}  \frac{\zeta^{2\alpha/\pi}}{[\zeta^2-a^2(t+\tau) ]^{\alpha/\pi} }\,d\zeta + a(t),
\end{equation}
which up to the first order in the parameter $\alpha$ becomes
\begin{equation}
 g_t \approx \int_{a(t+\tau)}^{g_{t+\tau}} \left\{1+ \frac{\alpha}{\pi}\left[2\ln\zeta-\ln(\zeta^2-a^2(t+\tau))\right] \right\} d\zeta + a(t),
\end{equation}
After performing the integral above and taking the limit $\tau \to 0$,  one readily obtains 
the following generalized Loewner equation:
\begin{equation}
 \dot{g}_t(z)=d(t)\left\{[g_{t} +a(t) ]\ln[g_{t}+a(t)]
 + [g_{t} -a(t) ]\ln[g_{t}-a(t)] 
 - 2g_t\ln g_t  \right\},
\label{eq:14}
\end{equation}
together with the governing equation for the logarithmic singularity $a(t)$ 
\begin{equation}
 \dot{a}(t) = (\ln4)  d(t) a(t),
 \label{eq:15}
\end{equation}
where the growth factor $d(t)$ is now given by
\begin{equation}
d(t)= \lim_{\tau \to 0}\frac{\alpha}{\pi \tau}.
\end{equation}
Equation (\ref{eq:15}) can be solved exactly to yield $a(t) = a_04^{\int_{0}^td(t')t'}$. As explained before, we can set $d(t)=1$ by rescaling the time coordinate, in which case we have $a(t) = a_04^t$. Recall also that the Loewner equation (\ref{eq:14}) must be supplemented with the initial condition $g_{0}(z)=z$.

\begin{figure}[t]
\begin{center}
\includegraphics[width=0.6\textwidth]{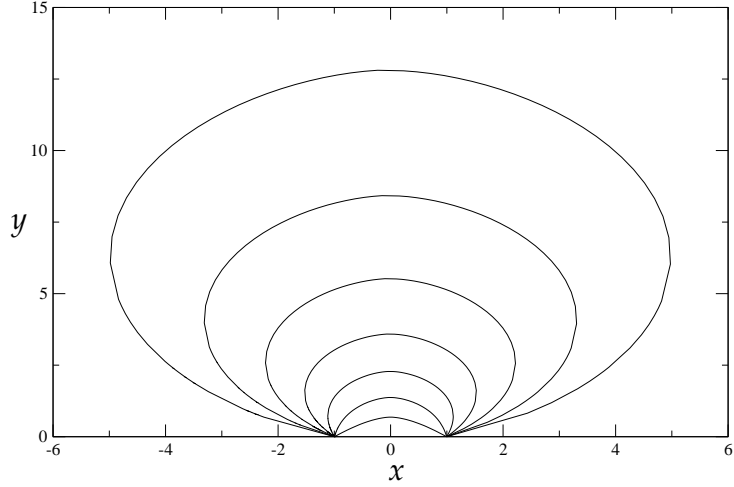}
\end{center}
\caption{\label{Fig. 5.} Loewner evolution for a bubble growing in the upper half-plane for $d(t)=1$. The bubble starts at $t=0$ from the interval $[-1,1]$ on the real axis. The different curves represent the interface at time intervals of $\Delta t=0.3$. }
\end{figure}

In Fig.~\ref{Fig. 5.}  we show numerical solutions of (\ref{eq:14}) with $a_{0}=1$ for various final times $t$, starting from $t=0.3$ up  $t=2.1$ with  a time separation of  $\Delta t =0.3$ between successive curves. To generate the curves shown in Fig.~\ref{Fig. 5.} we used the numerical scheme described in \cite{kadanoff2}. More specifically, we  start with a `terminal condition' $g_t=w$, for $w\in [-a(t),a(t)]$, and integrate the Loewner equation (\ref{eq:14}) backwards in time, using a Runge-Kutta method of second order, to get the corresponding point $g_0$ on the interface.  (See also \cite{kennedy2} for a recent review on numerical integration of the Loewner equation.) From Fig.~\ref{Fig. 5.} one sees that the bubble initially grows somewhat slowly and then rapidly expands and tends to occupy the whole plane for large times. In fact, one can show that  (for $d=1$) the tip velocity grows exponentially with time as $t\to\infty$. It is possible to modify the growth factor $d(t)$ so to have the tip velocity related to the gradient of the field $\phi(x,y)$, as discussed in Sec.~\ref{sec:2}, but this does not change the interface shapes and only alters the time scale of the bubble evolution.

\section{Conclusions}
\label{sec:4}

 We have presented a novel method to derive the Loewner equation for  Laplacian growth problems. The method is based on the Schwarz-Christoffel (SC) transformation and consists of expanding the integrand of the SC formula in the appropriate infinitesimal parameter, performing the relevant integrals, and then taking the limit in which the infinitesimal parameter goes to zero. Our method is able to reproduce the Loewner evolution  for the problem of slit-like fingers  in both the half-plane (Sec.~\ref{sec:2}) and the channel geometry (not shown here). Furthermore, the method can be extended to treat  more complicated growth problems, so long as the growth dynamics in the complex-potential plane  can be specified in terms of a polygonal boundary, in which case  the Schwarz-Christoffel transformation can be used. 
We note however that the requirement that the growth rule be formulated in terms of a polygonal curve is not as restrictive as it seems, for any simple curve can in principle be approximated by a piecewise linear function. [Such more general growth models are currently under investigation.]
 In particular, we obtained the Loewner equation for a novel situation in which a bubble grows from a segment of the real line into the upper half-plane. 
Although in this case we refer to the evolving interface as a growing `bubble,' in contrast to the slit-like `fingers' of Sec.~\ref{sec:2}, this terminology should not be taken literally. Depending on the physical problem at hand,  such growing interface may represent, say,  an expanding front  in combustion experiments or in electrochemical deposition \cite{poloneses}.  Of course, further work is necessary to relate more directly the growth models discussed here  to experiments.
 
\section*{Acknowledgments}

This
work was supported in part by the Brazilian agencies FINEP, CNPq, and FACEPE and 
by the special programs PRONEX and CT-PETRO. 




\medskip
\section*{References}
\smallskip

\end{document}